\documentclass[10pt]{article}

\usepackage[utf8]{inputenc}
\usepackage[T1]{fontenc}
\usepackage{amsmath,amssymb,graphicx}
\usepackage{subcaption}
\usepackage{algorithm}
\usepackage{algorithmicx}
\usepackage{natbib}
\usepackage{hyperref}
\usepackage{indentfirst}
\usepackage[a4paper,left=1.5cm,right=1.5cm,top=2cm,bottom=2cm]{geometry}

\usepackage{amssymb,amsthm,amsmath}
\usepackage{xcolor,paralist,hyperref,titlesec,fancyhdr,etoolbox}
\usepackage{lineno}

\titleformat{\section}[display]{\normalfont\huge\bfseries\centering}{}{0pt}{\Large}
\titlespacing*{\section}{0pt}{0ex}{0ex}

\begin{document}

\title{Casimir Arc--Plate Geometry: Computational Analysis of Thickness Constraints for Gold and Silver Nanomembranes in MEMS Applications}

\author{
Anna-Maria Alexandrova$^1$, 
Jesus Valdiviezo$^{2}$, 
\\
$^1$American College of Sofia, Sofia, Bulgaria \\
$^2$Sección Química, Pontificia Universidad Católica del Perú, Lima, Perú\\
}

\date{}

\maketitle

\renewcommand{\abstractname}{}
\begin{abstract}
A theoretical analysis of the Casimir interaction between an arc and plate is conducted, which remains unexplored despite its relevance to Micro-Electro-Mechanical Systems (MEMS) fabrication. The configuration consists of a rigid finite plate and a flexible curved nanomembrane, with radius $100 \mu m$, initially concave toward the rigid plate. The maximum thickness is evaluated for which the nanomembrane undergoes a change in curvature: from concave to convex with respect to the plate, due to the Casimir interaction. The Casimir energy for a curved surface is derived using the Proximity Force Approximation (PFA) with next-to-leading-order (NTLO) corrections. Kirchhoff-Love theory for a thin isotropic plate of constant thickness is used to estimate the bending energy. Material-dependent effects on the Casimir interaction are evaluated by comparing Au and Ag plates. The maximum thickness is derived where $U_{Casimir}>U_{bending}$ for distances in the range of $0.1-1 \mu m$. Results show curvature reversal occurs for nanomembranes with nanoscale thicknesses at the studied distances. Silver nanomembranes tolerate greater thickness than gold nanomembranes due to material-dependent properties. Comparison between NTLO-corrected PFA and perturbative PFA confirms the accuracy of the NTLO approach. The Casimir arc-to-plate geometry in MEMS enables Casimir-based actuation, enhances devices reliability, and prevents stiction. These findings provide thickness constraints for MEMS design and performance, accounting for the Casimir force.
\end{abstract}

\section{Introduction}
In 1947, in “The Influence of Retardation on the London-van der Waals Forces,” H. B. G. Casimir and D. Polder observed the London-van der Waals interaction of two neutral atoms. Their initial approach involved the use of second-order perturbation theory, considering the interaction of a neutral atom with a perfectly conducting wall, to measure the van der Waals force at increased separations \citep{casimir_influence_1948}. Further, Casimir treated the same problem in terms of vacuum; he calculated the zero-point energy of the electromagnetic field in the presence of molecules and compared it to the corresponding vacuum energy. Finally, to simplify this setup, Casimir reviewed the same interaction between two conductive plates rather than between molecules \citep{jaffe_casimir_2005}. This observation led directly to the prediction of the Casimir effect. 

\subsection{The Casimir Effect}

The Casimir effect is observed in the attractive force between two perfectly conducting, uncharged parallel plates, measured as in equation~\ref{casimir_parallel_equation} for distance $d$. The speed of light, $c$, transforms the electromagnetic mode wavelength into a frequency, which is further converted to energy by the reduced Planck constant. The negative sign represents the attractive force (later we will observe the conditions under which the Casimir force can act as a repulsive force) \citep{lamoreaux_casimir_2005}.

\begin{linenomath}
\begin{equation}
    \frac{F(d)}{A} = - \frac{\pi^2 \hbar c}{240 d^4}
    \label{casimir_parallel_equation}
\end{equation}
\end{linenomath}

 This effect is a consequence of the quantum vacuum: the vacuum fluctuations in the space between the plate and the outside space cause a pressure difference. In his calculations, Casimir proved that even at any finite volume, the zero-point energy of an electromagnetic vacuum is infinite, and changes in the energy can be identified. Specifically, this observation estimates the potential energy required to reduce the separation between the two plates from $d \rightarrow \infty$ , to distance $d$. This derivation fundamentally corresponds to the Casimir force \citep{milonni_quantum_1994}. The complete derivation can be reviewed from numerous sources \citep{lamoreaux_casimir_2005}, \citep{milonni_quantum_1994}.

\subsection{Geometry}

Although the derivation of the Casimir effect describes the interaction between two parallel plates, a variety of different geometries have arisen in the experimental setups. Configurations, such as plate-to-sphere \citep{lamoreaux_demonstration_1997}, sphere-to-sphere \citep{garrett_measurement_2018}, and plate-to-cylinder \citep{emig_casimir_2006}, have been experimentally evaluated. Owing to the challenge of keeping the geometry of the two plates parallel, many of the later performed experiments experienced a change in geometry, namely, the sphere-to-plate geometry. However, the vacuum energy estimate is geometry-dependent. Thus, the variation in the geometry of one of the two surfaces modifies the electromagnetic modes, resulting in varied energy differences. To express the Casimir effect for a curved surface, the proximity force approximation (PFA) method approximates the surface as a set of infinitesimal parallel plates, calculates the Casimir force for each pair of plates at the respective  distance, and integrates the contributions of all pairs to obtain the total force. For separation $d$ and curvature $R$, $d<<R$ must be true \citep{fosco_proximity_2011}. The simplified formula for the sphere-to-plate geometry, acquired from the PFA method, is described as in equation~\ref{casimir_force_sphere_equation}, where $R$ is the radius of the sphere and the condition $d<<R$ is met.

\begin{linenomath}
\begin{equation}
    \frac{F(d)}{A} = - \frac{\pi^3 \hbar c}{360} \frac{R}{d^3}
    \label{casimir_force_sphere_equation}
\end{equation}
\end{linenomath}

Theoretical analysis involves the geometries of a cylinder to a plate, parabolic cylinder to a plate, paraboloid to a plate, and sinusoidal corrugation to a plate \citep{fosco_proximity_2011}. However, PFA treats the locally flat regions as parallel to the opposing surface, and as a result it is considered an uncontrolled approximation. Next to leading order (NTLO) corrections are adapted to consider the effects of a curved geometry: a derivative expansion of the geometrical profile of the surface is considered up to the second term of the expansion \citep{fosco_proximity_2011}. Moreover, the study by \citep{fosco_casimir_2024} advanced corrections for finite temperature, two smooth surfaces, and various other cases. 

Unlike the sphere-plate geometry commonly analyzed in literature, including Ref.~\citep{stange_building_2019}, the arc--plate configuration considered in this work exhibits a spatially varying slope and a nonlinear variation of the local separation along the nanomembrane. In a sphere--plate geometry the surface slope changes smoothly and remains nearly constant over the interaction region, whereas for an arc geometry the slope can vary more rapidly, particularly near regions where the nanomembrane comes closest to the plate. As a result, the leading-order PFA alone may not fully capture the geometric dependence of the Casimir interaction. For this reason, the thickness estimates could benefit from NTLO corrections to ensure a consistent description of the Casimir energy in the arc--plate geometry, which is described later in our study.

\subsection{MEMS}

Owing to the microscale and magnitude of force compared to the scale, the Casimir effect has been applied in micro-electro-mechanical systems (MEMS) and sensors. The Casimir effect serves in quantum metrology, the approach of measuring parameters with increased accuracy using quantum phenomena. Quantum metrology has important applications in the detection of weak magnetic fields. The intended use of the Casimir effect in such devices, specifically magnetic sensors, utilized to assess brain and cardiac activity, cancerous tumors, and cosmic monitoring, is discussed in \citep{javor_analysis_2021}. In these systems, the Casimir effect augments the measurement resolution. Moreover, the study by \citep{stange_building_2019} effectively embedded a Casimir geometry on a commercial MEMS inertial sensor and measured the Casimir force. Monitoring the Casimir effect using a commercial MEMS device suggests its implementation as an engineering instrument in commercial MEMS. Furthermore, \citep{javor_zeptometer_2022} modeled a quantum displacement amplifier that converts DC to a much larger amplitude and frequency change in the AC. The Casimir effect is used to extract energy from the vacuum and supply it to a mechanical parametric amplifier. Combining the quantum-mechanical Casimir effect and MEMS results in a quantum metrology device, which can be applied to obtain zeptometer-level precision measurements.  

However, the major challenge in the application of the Casimir force in MEMS is stiction, referred to as Casimir pull-in, which can lead to device failure. In \citep{javor_analysis_2021}, the authors not only amplified the signals up to 10,000 times using the Casimir effect, but also prevented Casimir pull-in through time-delay-based parametric amplification. This is achieved by modulating the MEMS magnet oscillation parameter at specific frequencies and intentionally timing the oscillations to avoid collapse while achieving amplification. This displacement modulation counteracts the Casimir force, thereby ensuring that the components do not collapse. 

While prior theoretical studies have focused on plate-to-sphere or plate-to-cylinder configurations, arc-to-plate geometry, particularly for gold and silver nanomembranes, remains unexplored despite its practical relevance for MEMS applications. This study introduces a theoretical estimate to determine the maximum nanomembrane thickness that allows the arc to undergo curvature reversal under the Casimir force, enabling new insights for MEMS actuation. Additionally, these calculations provide a lower-bound constraint that can be applied in microfabrication to prevent stiction, demonstrating its practical use in improving the reliability of MEMS.

\section{Methods}

\subsection{Physical Interpretation and Experimental Methods of the Casimir Effect}
\subsubsection{The Quantum Vacuum}

Casimir recognized quantum electrodynamics (QED), where a vacuum has a lower energy limit for the system to be conserved, known as zero-point energy. Zero-point energy is essentially the sum of all possible frequencies, which explains its infinite magnitude. This is expressed in Peter W. Milonni’s “The Quantum Vacuum: An Introduction to Quantum Electrodynamics”, where the energy  $\sum_{k\lambda} \frac{1}{2} \hbar w_k$  is infinite \citep{milonni_quantum_1994}.

 To understand the concept of zero-point energy, one should recognize Heisenberg’s uncertainty principle, which is a necessary condition that combines three interrelated uncertainty relations for position and momentum \citep{milonni_quantum_1994}. First, the uncertainty principle constrains the widths of the position and momentum probability distribution in any quantum state; thus, when the position distribution is localized, the momentum distribution must be larger in the width area, and vice versa. Second, a disturbance uncertainty appears: there is a trade-off between measurements of the position and momentum; precisely measuring one causes a disturbance in the other. Third, owing to the non-commutativity of the position and momentum in the Heisenberg uncertainty bound, the constraints imposed by the joint approximation prevent the simultaneous determination of both quantities \citep{busch_heisenbergs_2007}. All three conditions describe Heisenberg's uncertainty, where the zero-point energy is explained specifically by the energy-time relation (equation~\ref{heisenber_et_equation}), which is a specific form of the generalized Heisenberg inequality (equation~\ref{heisenber_main_equation}). To explain vacuum, an inverse scenario is observed: if the system has energy, s.t. $\Delta E =0$ , the time would have to reach infinity. Considering a finite-time system, the energy value remains uncertain. Therefore, energy there must be present in the vacuum. In equation~\ref{heisenber_et_equation} $\Delta E$ is the uncertainty in energy, $\Delta t$ is the uncertainty in time, and $\hbar$ is the reduced Planck constant. In equation~\ref{heisenber_main_equation} $\sigma_a$ and $\sigma_b$ are the standard deviations (uncertainties) of the observables $A$ and $B$, $[A,B] = AB - BA$ is their commutator, $\langle [A,B] \rangle$ denotes its expectation value, and $i$ is the imaginary unit.

\begin{linenomath}
\begin{equation}
     \Delta E \Delta t \gtrsim \frac{\hbar}{2} 
     \label{heisenber_et_equation}
\end{equation}
\end{linenomath}

\begin{linenomath}
     \begin{equation}
     \sigma_a \sigma_b \geq \left| \frac{1}{2i} \, \langle [A, B] \rangle \right|^2
\label{heisenber_main_equation}
 \end{equation}
 \end{linenomath}

To maintain the energy-time uncertainty relation, quantum vacuum fluctuations arise simply as random, temporary changes in energy. The appearance and disappearance of virtual particles serve as mathematical representations of disturbances in the field. 
As a result of the vacuum field, many phenomena can be observed, some of which are the Lamb shift, atoms spontaneously emitting photons, and Casimir force \citep{milonni_quantum_1994}.   

\subsubsection{Boundary Conditions}

In “On the attraction between two perfectly conducting plates”(1948) Casimir shows that the electromagnetic field experiences changes to satisfy the boundary conditions introduced by the plates \citep{casimir_attraction_nodate}. Specifically, a pressure difference was observed, which is the cause of the assumption that the plates have infinite conductivity. The perfect conductivity of the plates creates an electromagnetic field within the conductor, $E=0$. To satisfy Maxwell’s boundary conditions, the tangential component vanishes \citep{jackson_classical_2009}. Mathematically, this corresponds to the Dirichlet boundary condition, which requires a field to be fixed at zero in the plates. Moreover, the enclosed space between the plates allows only wavelengths of certain $\lambda$ to propagate in the field. For $\lambda$, equation~\ref{lambda_2d_n_equation} expresses the boundary conditions that should be satisfied, where, $\lambda$ is the allowed wavelength of the field between the plates, $d$ is the distance between the plates, and $n$ is a positive integer ($n = 1, 2, 3, ...$) representing the mode number.

\begin{linenomath}
\begin{equation}
    \lambda = \frac{2d}{n} 
    \label{lambda_2d_n_equation}
\end{equation}
\end{linenomath}

\subsubsection{Repulsive Casimir effect}

Although the Casimir force is attractive at the origin, a repulsive Casimir force was measured in \citep{munday_measured_2009} when solid materials were immersed in a fluid. Munday, Capasso and Parsegian considered the dielectric response functions, $\epsilon$, of the materials. They observed that in a configuration of $\epsilon_1>\epsilon_3> \epsilon_2$ , where $\epsilon_3$ represents the medium and $\epsilon_1$ and $\epsilon_2$ the materials, the force is repulsive. These findings suggest applications such as quantum levitation, allowing the development of ultra-low friction devices and sensors \citep{munday_measured_2009}. In addition, the repulsive Casimir force was observed in \citep{zhao_stable_2019} in a system consisting of a gold surface and a Teflon coated gold surface, immersed in ethanol.

Other methods for obtaining a repulsive Casimir force involve modifications of the geometry, such as the proposed glide-symmetric geometry \citep{rodriguez_repulsive_2008}. We observed that a change in geometry altered the allowed field modes. The repulsive Casimir force case is usually achieved by introducing an asymmetry between the two surfaces. Thus, both attractive and repulsive Casimir forces can be applied to MEMS. 

\subsubsection{Experiments and Materials}
In experiments that estimate the Casimir force, researchers seek materials with high conductivity yet minimal surface roughness to obtain precise measurements.

With a conductivity value of $4.5 \times 10^7 S/m$, gold is the most commonly used material because of its ability to be modified to achieve minimal roughness. Methods involve template stripping as suggested in \citep{ederth_template-stripped_2000}, achieving a root mean square roughness of less than $0.4$ nm, which is crucial for the accuracy of measurements at nanoscale distances. Moreover, \citep{banishev_modifying_2012} distinguished the difference in the observed force when both bodies were made of Au, compared to when one body was made of Au and the other of Si. The force between the Au sphere and Si plate geometry with respect to the Au sphere and Au plate configuration was represented in a ratio of $0.74$ at separations of $100$ nm and $0.63$ at separations of $200$ nm , as recognized in \citep{chen_investigation_2005}. 

Silver has an even higher conductivity value of $6.3 \times 10^7 S/m$, which should theoretically increase the magnitude of the Casimir force. However, owing to overtime oxidation, the surface roughness increases and, at the nanoscale, this will affect the accuracy of measurements, making gold preferred in experimental setups.
The oxidation rate of silver increases with a decrease in the size of silver particles, notably for diameters $d_p \leq 6$ nm \citep{chaparro_oxidation_2023}.
Furthermore, the oxidation process is shape-dependent, as described in \citep{chaparro_shape-dependent_2024}, where the oxidation process of silver nanostructures is observed, specifically for spheres, cubes, disks, cylinders, triangles, and pyramids. The nanospheres had the lowest level of oxidation among all the geometries tested \citep{chaparro_shape-dependent_2024}. Selecting specific silver nanostructures according to their oxidation characteristics can improve the design of MEMS and NEMS. Thus, the material properties affect the accuracy of the theoretical results compared to the experimental evaluation. 

The Casimir force is observed as dominant in systems at the microscale and has been further experimentally verified. The Casimir force is observed as dominant in systems at the microscale and has been further experimentally verified. Lamoreaux conducted the first experimental measurement of the Casimir force between a flat plate and a sphere using a torsion pendulum for separations of $0.6$ to $6 \mu m$. Both components of the composed geometry were coated with a $0.5 \mu m$ Cu layer on each side, and further $0.5 \mu m$ Au was evaporated on the faces undergoing attraction. The theoretical results agreed with the experimental measurements within an error of $5\%$ \citep{lamoreaux_demonstration_1997}. Another common method involves atomic force microscopy (AFM), owing to the high accuracy of measurements that it allows. U. Mohideen and A. Roy first implemented the AFM method for a plate-to-sphere geometry with a separation range of $0.1$ to $0.9 \mu m$, and obtained an accuracy of $1\%$ \citep{mohideen_precision_1998}.

\subsection{Theoretical Framework}

 The aim of this study is to observe the critical thickness that allows the arc to reverse its curvature from concave to convex with respect to the plate. The estimates are observed for two different material choices. The objective of this theoretical analysis is to simulate an experimental evaluation: the distance range, radius of curvature of the arc, and materials chosen s.t. can be replicated using micromachinery and observed experimentally. For the purposes of this study, we established a theoretical setup involving the geometry, choice of materials, and distance range, as described in the following section.

\begin{figure}[h]
 
    \centering
    \begin{subfigure}[b]{0.45\textwidth}
    \includegraphics[width=1\textwidth]{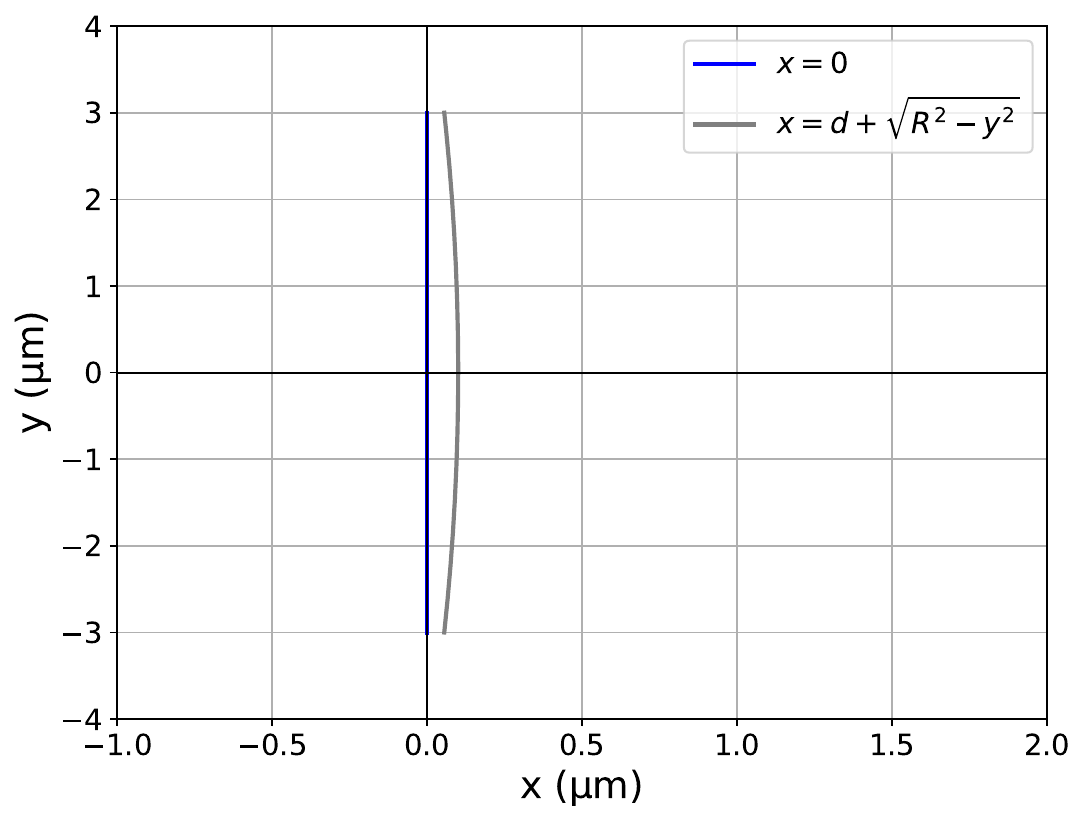}
    \caption{}
        \label{fig:arc_to_plate_d_01}
    \end{subfigure}
    \hfill
    \begin{subfigure}[b]{0.45\textwidth}
    \includegraphics[width=1\textwidth]{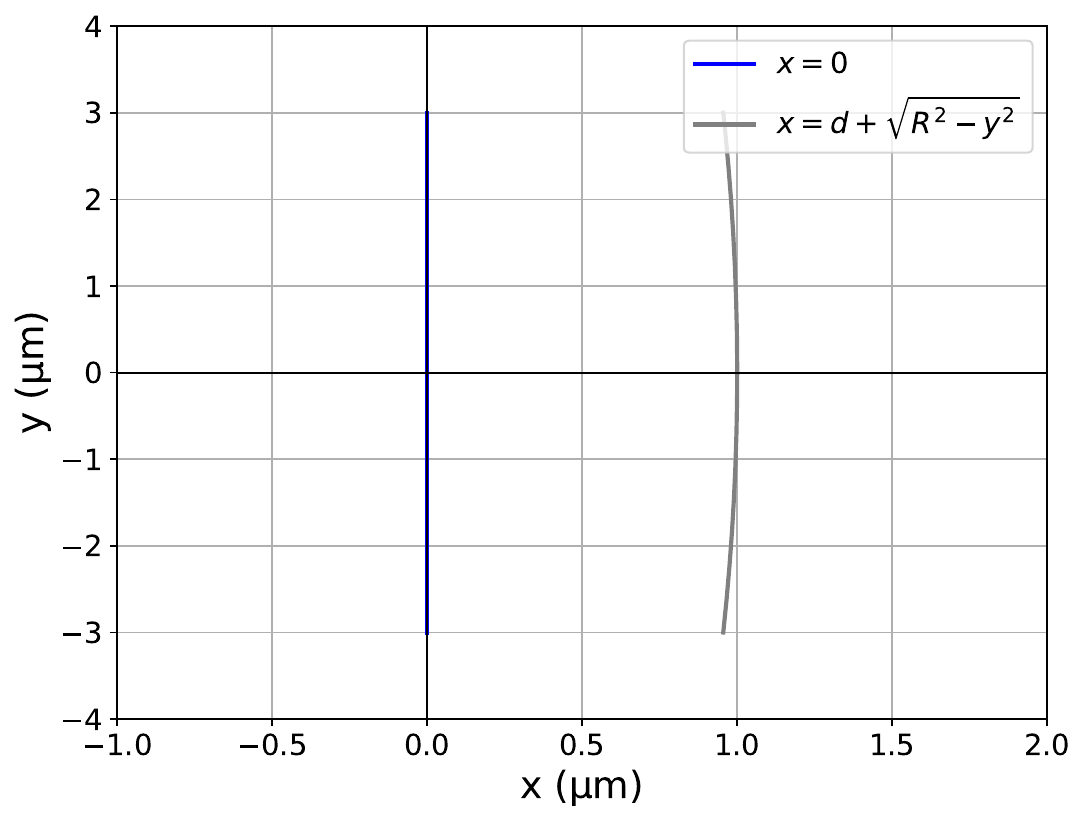}
    \caption{}
        \label{fig:arc_to_plate_d_1}
    \end{subfigure}
\caption{\raggedright {Arc-to-plate geometry at distance (a) $0.1 \mu m$ (b) $1 \mu m$}.}
    \label{fig:arc_geometry}
\end{figure}

First, a Casimir geometry consisting of an arc and a plate was introduced. The arc is essentially a finite curved nanomembrane, with length which we approximate to $6\mu m$ (due to the large radius of $R=100 \mu m),$ concave to a finite plate with length 6 $\mu m$ (figure~\ref{fig:arc_to_plate_d_01}). Both surfaces were fixed so that the Casimir interaction could not bring them together. 
The nanomembrane is modeled as a thin elastic arc with clamped endpoints, as a representation of a membrane fixed to a rigid MEMS frame. Clamped boundary conditions are assumed at the ends of the arc, which constrains both displacement and rotation. These assumptions are consistent with typical MEMS design \citep{klimchitskaya_casimir_2024}. Future work could extend these results through formal stability analysis.
Figure~\ref{fig:arc_to_plate_d_01} represents the arc-plate geometry at a separation of $0.1 \mu m$ of the two surfaces, and figure~\ref{fig:arc_to_plate_d_1} shows the arc-plate geometry at a distance $1 \mu m$. Reference~\citep{gies_casimir_2006} shows that Casimir edge effects for finite plates are negligible when the finite extent of the plate in the direction parallel to its surface is sufficiently small. In this study, we consider the plate and arc to be sufficiently small in the direction parallel to their surfaces to satisfy this condition, allowing edge effects to be safely neglected throughout our analysis.

The arc represents a segment of a circle, and is described by equation~\ref{geometry_arc_equation}. Here, the distance $d$ is defined s.t. $d = d_0 -R$ , where $d_0$ is the distance between the two surfaces in the chosen distance range, $R$ is the radius of the arc, and each $(x, y)$ represents a pair corresponding to a specific point on the arc, with $y$ chosen along the vertical span and $x$ determined by the circular shape.

\begin{linenomath}
\begin{equation}
x = d+ \sqrt{R^2-y^2}
\label{geometry_arc_equation}
\end{equation}
\end{linenomath}

The distance range between the arc and the plate, considered from the farthest point of the arc, is described as $d \in [0.1 - 1] \mu m$. Although the Casimir force increases with decreasing distance, most experiments operate in the micro- or sub-micro range, as in \citep{mohideen_precision_1998}, \citep{bimonte_measurement_2021}. Decreasing the distance to the nanoscale introduces challenges such as effects due to surface roughness and inability to achieve precision as the Casimir force becomes uncontrollably strong.

\begin{figure}[h!]
    \centering
    \includegraphics[width=1\textwidth]{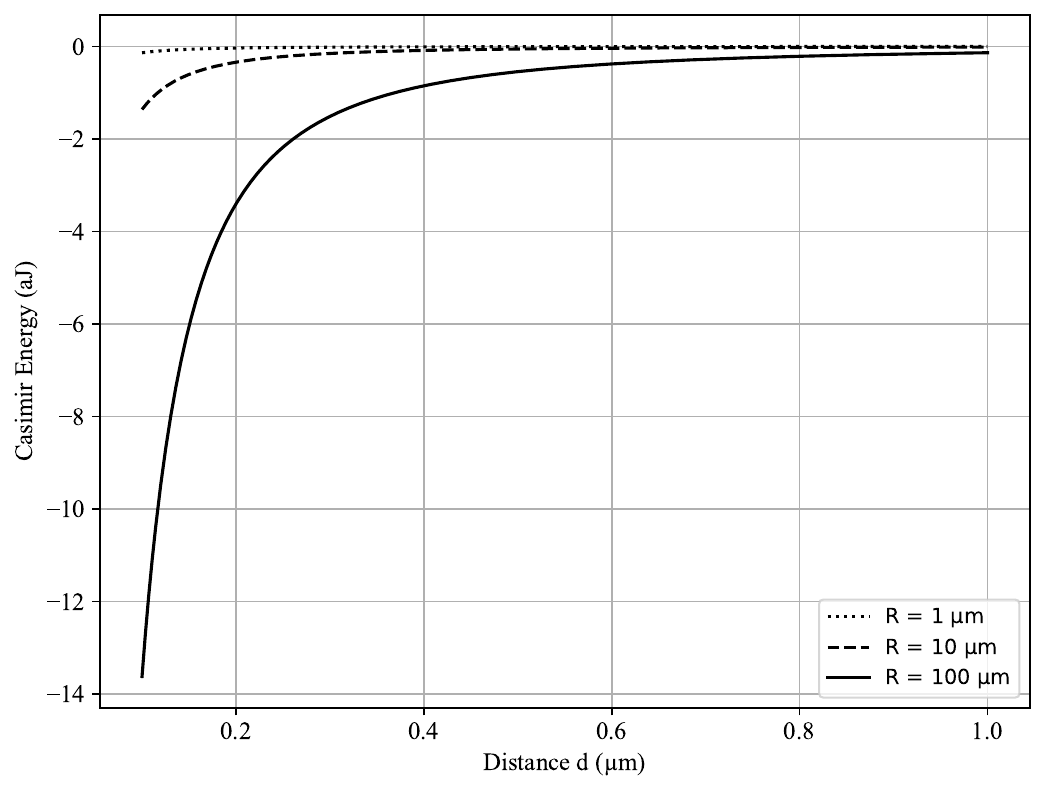}
    \caption{{Casimir energy vs distance for various radii in a sphere-to-plate geometry.}}
    \label{fig:Energy_vs_Radii}
\end{figure}

In order to satisfy the conditions which PFA implies, the radius is chosen to minimize the error, that is $d/R<<1$. A radius of $100 \mu m$ is employed. Equation~\ref{casimir_force_sphere_equation} implies the direct proportionality of the radius and Casimir force, suggesting an increasing force magnitude with an increasing radius value, as shown in figure~\ref{fig:Energy_vs_Radii}. Figure~\ref{fig:Energy_vs_Radii} shows the Casimir energy for different values of $R (R=1, R=10, R=100 \mu m$) in the range of $0.1$ to $1 \mu m$. $R=100 \mu m$ as observed contributes to the highest Casimir energy.

The material choice is explained in Section 2.1.4. "Experiments and Materials" of this paper. The values for the Poisson's ratio and Young's modulus were taken into account in the case of a thin film for both gold and silver.

\subsection{Conditions}

The conditions under which this model operates are introduced. 

To satisfy the PFA condition, $d<<R$ or $d/R<<1$ must be satisfied.

Applying Kirchhoff-Love's thin plate theory assumes that the plates are of uniform thickness and isotropic material. Moreover, the thin-plate assumption requires a plate of thickness $t$ that satisfies the conditions described in equations~\ref{kirchhoff_love_c1},~\ref{kicrhoff_love_c2}, and ~\ref{kirchhoff_love_c3} , where $a$ and $b$ represent the width and length, respectively.

\begin{linenomath}
\begin{equation}
{
   t<< a \text{ and } t<<b 
   \label{kirchhoff_love_c1}
} 
\end{equation}
\end{linenomath}

\begin{linenomath}
\begin{equation}
{
   t< \frac{1}{10}a \text{ for } a>b
   \label{kicrhoff_love_c2}
}
\end{equation}
\end{linenomath}

\begin{linenomath}
\begin{equation}
{
   t< \frac{1}{10}b \text{ for } b>a
   \label{kirchhoff_love_c3}
} 
\end{equation}
\end{linenomath}

Finally, the strain in the $z$ direction is equal to $0$ , so that a simplification of the three-dimensional problem converts it into a two-dimensional one.

\subsection{Procedure}

A two-dimensional model of arc-plate geometry was considered. 

In classical plate and shell theory, curvature reversal corresponds to a buckling-type instability whose critical condition depends on boundary constraints and in-plane stresses. A full buckling analysis under Casimir loading would require solving nonlinear equilibrium equations and identifying bifurcation points for curved beams or shells, typically involving eigenvalue problems and numerical methods \citep{emam_review_2022}. Such an analysis is beyond the scope of our study and is left for future investigation.

In this study, curvature reversal is estimated by comparing the Casimir-induced energy to the elastic bending energy. The results provide an upper bound on the critical thickness for instability. This energy-based comparison does not constitute a full buckling analysis, but serves as a conservative criterion for the onset of curvature instability.

To evaluate the critical thickness, the Casimir and bending energies were compared. 
The bending of the arc from concave to convex with respect to the plate occurs only when the Casimir energy overpowers the total energy required to bend the nanomembrane, as shown in equation~\ref{Ub_U_c_equation}, where $ U_{Casimir}$ represents the Casimir energy and $U_{Bending}$ represents the bending energy.

\begin{linenomath}
\begin{equation}
    U_{Casimir}> U_{Bending}
    \label{Ub_U_c_equation}
\end{equation}
\end{linenomath}

\subsubsection{Calculating Casimir Energy}

The PFA is used to obtain the Casimir energy, estimated as in equation~\ref{Upfa} where the total energy is obtained by integrating the local energy density \(E_{||}(d(r))\) (as expressed in equation~\ref{E||}) over the surface \(S\). The local energy density corresponds to two parallel plates separated by a distance $d(r)$, which varies along the curved geometry.

\begin{linenomath}
\begin{equation}
U_{PFA} = \int dS E_{||}(d(r))
\label{Upfa}
\end{equation}
\end{linenomath}

\begin{linenomath}
\begin{equation}
E_{||}(d) = - \frac{\pi \hbar c}{720 d^3}
\label{E||}
\end{equation}
\end{linenomath}

Fosco’s correction was implemented by applying a derivative expansion to the first two terms as described in equation~\ref{fosco} where $U_{DE}$ is the Casimir energy including the NTLO correction, $\hbar$ is the reduced Planck constant, $c$ is the speed of light in vacuum, $x_{||}$ denotes the coordinates parallel to the reference surface over which the integration is performed, $\psi(d)$ is the local separation, $d$, between the curved surface and the opposing plate (equation~\ref{geometry_equation}), and $(\partial \psi)^2$ represents the sum of the squared partial derivatives of $\psi$ with respect to the parallel coordinates, i.e., $(\partial \psi)^2 = (\partial \psi/\partial x)^2 + (\partial \psi/\partial y)^2$ \citep{fosco_proximity_2011}. 

The geometric function describes the arc, as expressed in equation~\ref{geometry_equation}. Restrictions were imposed, namely $[-3<y<3]$, to prevent contact between the two fixed surfaces. As described in \cite{fosco_proximity_2011}, the first term in the derivative expansion corresponds to PFA. Thus, if one approximates the second term as a small perturbation parameter, $\epsilon$, by perturbation theory, $U_{DE}$ reduces the expression to the PFA.

NTLO corrections serve to better approximate the Casimir energy for the arc geometry due to its geometrical properties, accounting for variations in the local slope where distances change nonlinearly. These corrections are derived using a derivative expansion, which is valid when the local surface curvature varies slowly compared to the separation distance. This condition is satisfied in our study for all distances \( d \in [0.1, 1]\,\mu\mathrm{m} \) with a curvature radius \( R = 100\,\mu\mathrm{m} \). This ensures that the NTLO correction remains controlled and physically meaningful. Experimental verification of the accuracy of PFA with NTLO corrections is left for future work.

\begin{linenomath}
\begin{equation}
    U_{DE} = - \frac{\pi^2 \hbar c}{1440} \int d^2 x_{||} \frac{1}{\psi^3}[1+\frac{2}{3}(\partial\psi)^2]
    \label{fosco}
\end{equation}
\end{linenomath}

\begin{linenomath}
\begin{equation}
    \psi(d) = x(d) = d + \sqrt{R^2 - y^2} 
    \label{geometry_equation}
\end{equation}
\end{linenomath}

\subsubsection{Calculating Bending Energy}

The bending energy, $U_{bending}$, is obtained using equation~\ref{u_bending_integral_equation}, where $u$ is the bending strain energy and $A$ is the area \citep{wierzbicki_tomasz_2081j16230j_nodate}. Because we obtain a numerical value for $u$, the equation simplifies to equation \ref{u_bending_l_equation}, where $A$ simplifies to $L$, the length of the arc. 
To evaluate the arc length, $L$ is defined in equation~\ref{arc_length_equation}, where $y_{min} = -3\mu m$ and $y_{max} = 3\mu m$. The length of the arc was $L = 6 \mu m$.

\begin{linenomath}
\begin{equation}
    U_{bending} = \int_A u  dA
    \label{u_bending_integral_equation}
\end{equation}
\end{linenomath}

\begin{linenomath}
\begin{equation}
    U_{bending} = u\times L 
    \label{u_bending_l_equation}
\end{equation}
\end{linenomath}

\begin{linenomath}
\begin{equation}
L = \int_{y_{\min}}^{y_{\max}} \sqrt{1 + \left(\frac{dx}{dy}\right)^2} \, dy
\label{arc_length_equation}
\end{equation}
\end{linenomath}

A breakdown of the bending energy equation~\ref{u_bending_integral_equation} involves the equation for the bending strain energy (equation~\ref{u_equation}) and the equation for the bending stiffness (equation~\ref{bending_stiffness_equation}). Equation~\ref{u_equation} describes the bending strain energy density per unit area, where $k$, the local curvature, estimated as $k =\frac{1}{R}$, $D$ is the bending stiffness as described in equation~\ref{bending_stiffness_equation}, where $E$ represents Young's modulus, $\upsilon$ is Poisson's ratio, and $t$ is the thickness of the plate. 

\begin{linenomath}
\begin{equation}
{
u = \frac{D}{2} {\left[ (k_{11} + k_{22})^2 - 2(1-\nu) \left( k_{11} k_{22} - k_{12}^2 \right) \right]}
\label{u_equation}
}
\end{equation}   
\end{linenomath}

\begin{linenomath}
\begin{equation}
    D = \frac{E t^3}{12(1 - \upsilon^2)}
    \label{bending_stiffness_equation}
\end{equation}
\end{linenomath}

Table~\ref{tab:materials} shows the values of $E$, Young's modulus, and $\upsilon$, Poisson's ratio, for thin gold and thin silver films \citep{noel_review_2016}, \citep{ruud_elastic_1991}. 

\begin{table}[h]
\caption{{Properties of gold and silver}: $E$, Young's modulus and $\upsilon$, Poisson's ratio.}
\centering
\begin{tabular}{|c|c|c|}
\hline
\textbf{} & \textbf{Gold (Au)} & \textbf{Silver (Ag)} \\
\hline
$E$ [GPa, $10^9$ Pa] & $97 \pm 10$ & 83.6 \\
\hline
$\upsilon$ & $0.421 \pm 0.06$ & 0.517 \\
\hline
\end{tabular}

\label{tab:materials}
\end{table}

To define the arc geometry, the curvature tensor is defined as follows:

\[
k_{12} =
\begin{bmatrix}

k_{11} & k_{12} \\
k_{21} & k_{22}

\end{bmatrix}
\]

\[
k_{12} =
\begin{bmatrix}
\frac{1}{R} & 0 \\
0 & 0
\end{bmatrix}
\]

Substituting into equation~\ref{u_equation}, the bending strain energy results as in equation~\ref{u_simplified_equation}, where a substitution for the value of $D$ should be made for both gold and silver.

\begin{linenomath}
\begin{equation}
    u = \frac{D}{2\times10^{-8}}
    \label{u_simplified_equation}
\end{equation}
\end{linenomath}

For $
U_{Casimir}> U_{Bending} $ the values of $U_{gold}$ and $U_{silver}$ were introduced as constants multiplied by the cubed thickness. $U_{const}$ represents the value at which both $U_{goldconst}$ and $U_{silverconst}$ are substituted with.

\[
\Rightarrow U_{Casimir}>U_{const} \times t^3
\]
The final equation, which isolates the critical thickness, results in the following:

\begin{linenomath}
\begin{equation}
    \left(\frac{U_{Casimir}}{U_{goldconst}}\right)^{1/3} > t
\end{equation} 
\end{linenomath}

Ultimately, this is the maximum thickness the nanomembrane can hold to shift its curvature. 
In the simulation, it was calculated by $(\frac{U_{Casimir}}{U_{goldconst}})^{1/3} = t$ for 1000 distances in the distance range $d \in[0.1 - 1] \mu m$.
It should be noted that, owing to the attractive force, a negative sign appears in front of the Casimir energy; thus, the absolute value of $U_{Casimir}$ is taken.

\subsection{Algorithm for Critical Thickness Evaluation}

The procedure for evaluating the critical thickness can be summarized in the following algorithm.

\begin{algorithm}[H]
\caption{Evaluation of Critical Thickness in Arc to Plate Geometry}
\begin{algorithmic}[1]

\Statex \textbf{1: Define Geometry and Parameters}
\State Introduce arc to plate geometry (figures 1 and 2).
\State Set distance range $d \in [0.1, 1] \,\mu m$.
\State Choose radius $R = 100 \,\mu m$ to satisfy $d/R \ll 1$.

\Statex \textbf{2: Calculate Casimir Energy}
\State Use PFA:
\[
U_{PFA} = \int dS \, E_{||}(d(r)), \quad E_{||}(d) = -\frac{\pi \hbar c}{720 d^3}
\]
\State Apply derivative expansion correction \cite{fosco_proximity_2011}:
\[
U_{DE} = - \frac{\pi^2 \hbar c}{1440} \int d^2 x_{||} \frac{1}{\psi^3} \left( 1 + \tfrac{2}{3}(\partial \psi)^2 \right)
\]

\Statex \textbf{3: Calculate Bending Energy}
\State Define bending stiffness for tested materials:
\[
D = \frac{E t^3}{12 (1-\nu^2)}
\]

\State Express bending energy:
\[
U_{bending} = \int_A u \, dA, \quad 
u = \frac{D}{2} \left[(k_{11} + k_{22})^2 - 2(1-\nu)(k_{11}k_{22}-k_{12}^2)\right]
\]
\State Substitute curvature tensor:
\[
k_{ij} =
\begin{bmatrix}
\frac{1}{R} & 0 \\ 
0 & 0
\end{bmatrix}
\]

\Statex \textbf{4: Compare Casimir and Bending Energy}
\State Critical condition:
\[
U_{Casimir} > U_{bending} \quad \Rightarrow \quad t < \left(\frac{U_{Casimir}}{U_{const}}\right)^{1/3}
\]

\Statex \textbf{5: Numerical Simulation}
\State Evaluate critical thickness for 1000 distance values in $d \in [0.1, 1] \,\mu m$.
\State Compare results for gold and silver films.

\end{algorithmic}
\end{algorithm}

\section{Results}

\subsection{Estimating the maximum thickness using PFA with NTLO corrections}

The results of this study include the estimated maximum thickness, allowing for the reversal of the curvature of the nanomembrane (from convex to concave with respect to the plate), calculated for an array of 1000 distances within the distance range $d \in [0.1 - 1] \mu m$ between the arc and plate. The maximum thicknesses of gold and silver were evaluated and compared. 

The results show that both silver and gold nanomembrane thickness values operate at the nanoscale level. The range of values is $t \in [10^{-9}$ \textendash $ 10^{-10}]$ m. However, the maximum thickness of Ag nanomembrane was slightly larger than that of the Au nanomembrane. This is due to the fact that the Casimir energy, as estimated in this paper, is solely geometry dependent; moreover, both Casimir setups for gold and silver are of exact geometry. Therefore, the difference in thickness between gold and silver is based entirely on the bending energy. Specifically, $U_{bending}$ of silver is less than $U_{bending}$ of gold. Thus, this accounts for the effects of the material properties. 
The distance and radius of curvature affect the Casimir energy. A predictable trend is observed, introduced by the general Casimir relationship: as the distance increases, the energy decreases; as a result, the thickness values become smaller. Moreover, because of the bending energy calculations, the thickness value followed the cube root of the Casimir energy. However, the relationship between the Casimir energy and the radius is proportional. Therefore, with increasing radius, the increased Casimir energy allows for a greater thickness value.

\begin{figure}[h]
 
    \centering
    \begin{subfigure}[b]{0.45\textwidth}
       \includegraphics[width=1\textwidth]{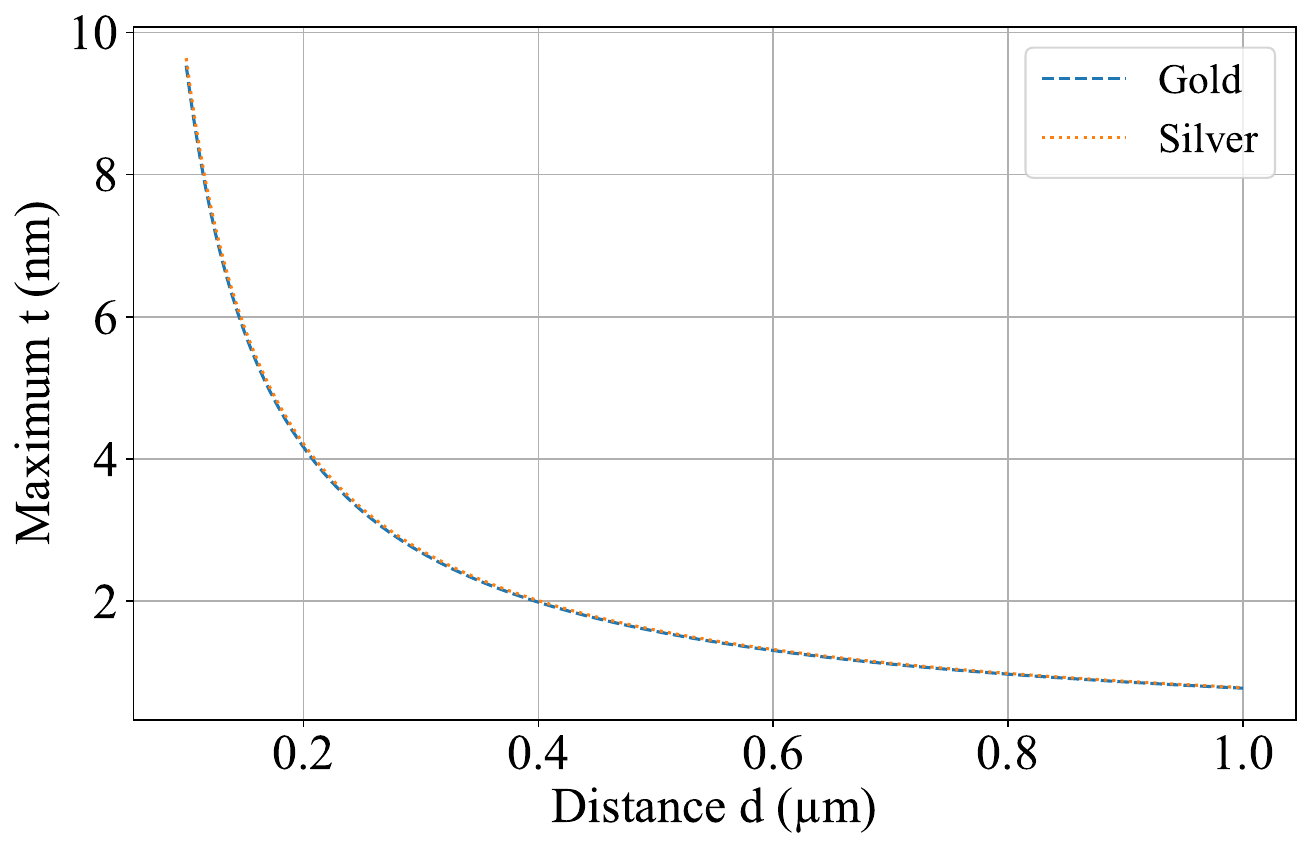}
    \caption{}
    \label{fig:graph_results_thickness}
    \end{subfigure}
    \hfill
    \begin{subfigure}[b]{0.45\textwidth}
     \includegraphics[width=1\textwidth]{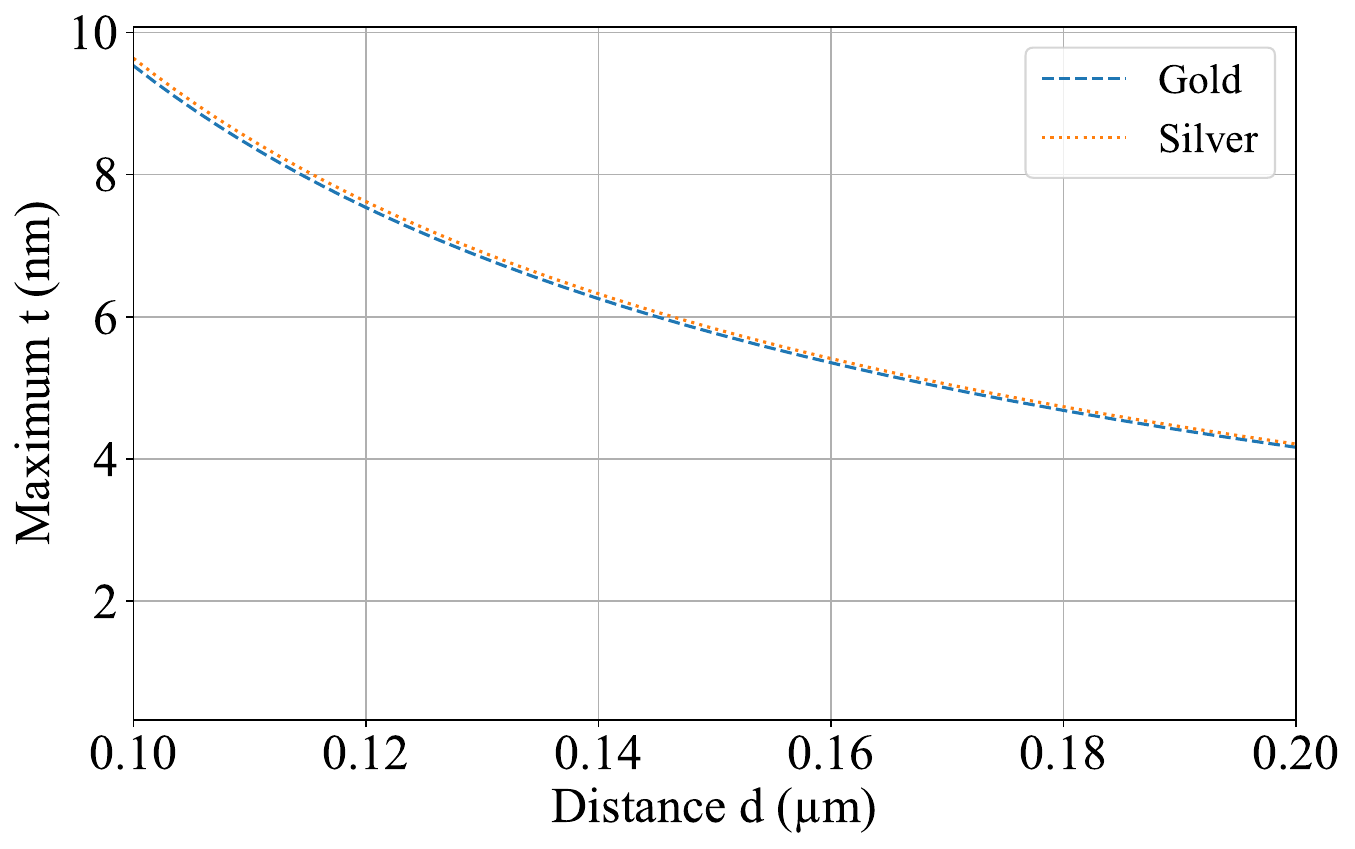}
    \caption{}
    \label{fig:graph_results_thickness_zoomed}
    \end{subfigure}
\caption{{\raggedright {Maximum thickness of Au and Ag nanomembranes: corrected PFA with NTLO} for the distance range of (a) $0.1 \le d \le 1 \mu m$.{(b) $0.1 \le d \le 0.2 \mu m$.}}}
    \label{fig:thickness_results}
\end{figure}

Tale~\ref{fig:table_results_thickness} summarizes 10 of the values for an increment $\Delta d = 0.1 \mu m$ between the distances for each silver and gold. Figure~\ref{fig:graph_results_thickness} shows the maximum thickness for all values of $d$, displaying separate curves for Au and Ag. Both materials are influenced by the Casimir force, which follows an inverse relationship with distance. Thus, the curve exhibits exponential decay: $t$ decreases logarithmically with increasing distance. Figure~\ref{fig:graph_results_thickness_zoomed} focuses on the distance range $0.1 - 0.2 \mu m$, where the thickness appears at its maximum values from the distance range.
    
\begin{table}[H]
    \centering
    \captionsetup{justification=centering, skip=0pt}
      \caption{{Maximum thickness of Au and Ag nanomembranes: corrected PFA with NTLO.}}
      \vspace{-2pt}
    \includegraphics[width=1\textwidth]{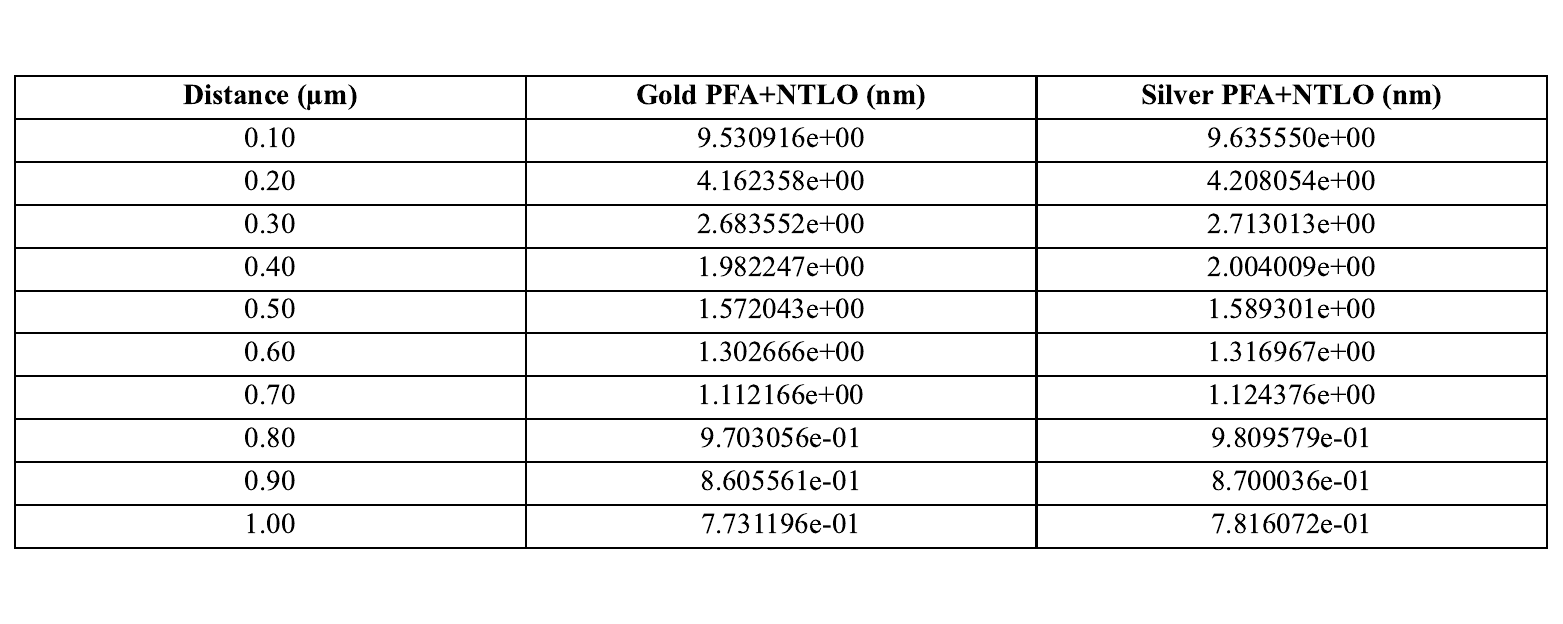}
    \label{fig:table_results_thickness}
\end{table}

\subsection{Comparison to PFA with $\epsilon = 0.1$ as the perturbation parameter}

Including higher-order corrections to the PFA contributes to a more realistic estimate of the results. However, our analysis shows that the higher-order correction is negligible, modifying the results only at the level of the fourth significant digit. 

\begin{table}[h!]
    \centering
     \captionsetup{justification=centering, skip=0pt}
    \caption{{{Maximum thickness of Au and Ag nanomembranes at selected distances:
PFA with $\epsilon= 0.1$ as the perturbation parameter.}}}
\vspace{-2pt}
    \includegraphics[width=1\textwidth]{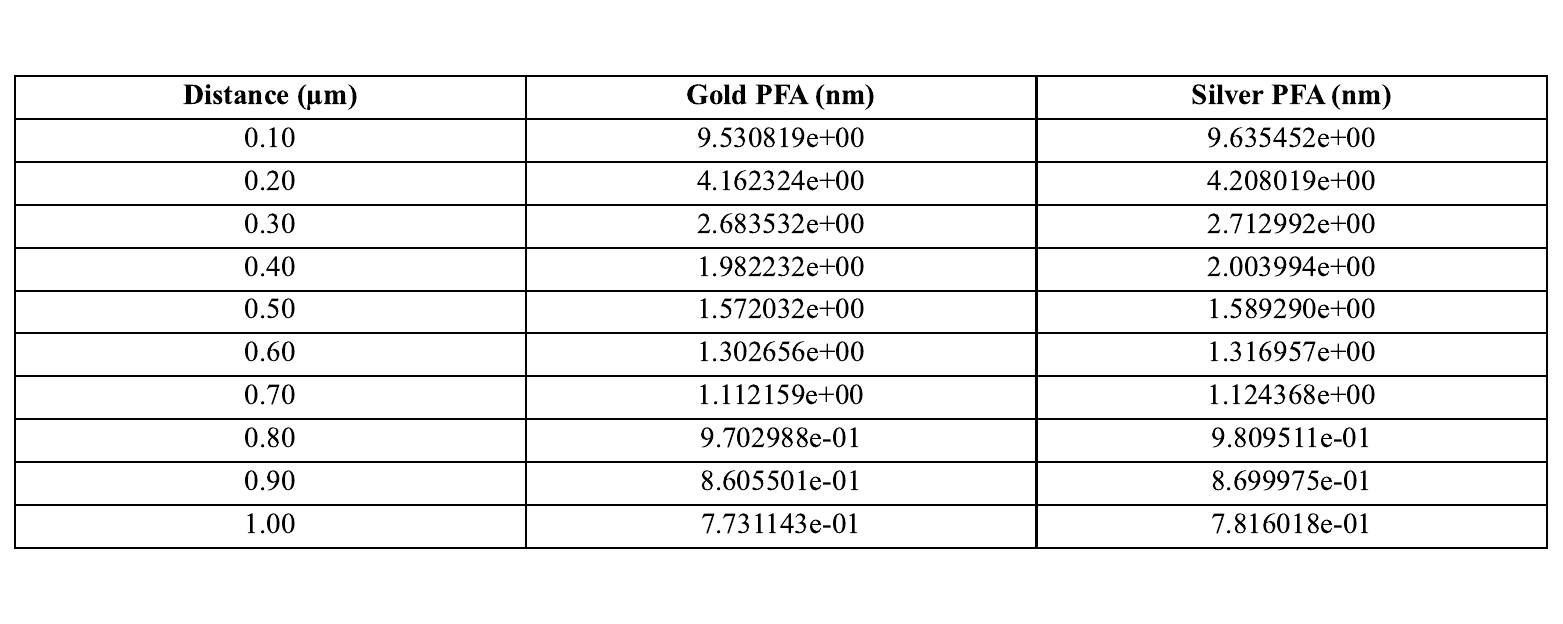}
    \label{fig:table_PFA_perturbation}
\end{table}

We quantify this estimate by the fractional deviation $\delta$, described as a function of $d$, which is obtained as in equation~\ref{error_equation}, where $t_{PFA}$ is the thickness value obtained from PFA with  perturbation parameter $\epsilon=0.1$, and $t_{PFA+NTLO}$ is the thickness value received from the PFA method with NTLO correction. 
\begin{linenomath}
\begin{equation}
   {\delta(d)}=\frac{|t_{PFA} - t_{PFA+NTLO}|}{t_{PFA+NTLO}}
    \label{error_equation}
\end{equation}
\end{linenomath}

The quantity defined in equation~\ref{error_equation} represents the fractional deviation between the leading-order PFA and the PFA with NTLO correction. Larger values indicate a stronger contribution from higher-order geometric corrections.

\begin{figure}[h]
 
    \centering
    \begin{subfigure}[b]{0.48\textwidth}
       \includegraphics[width=1\textwidth]{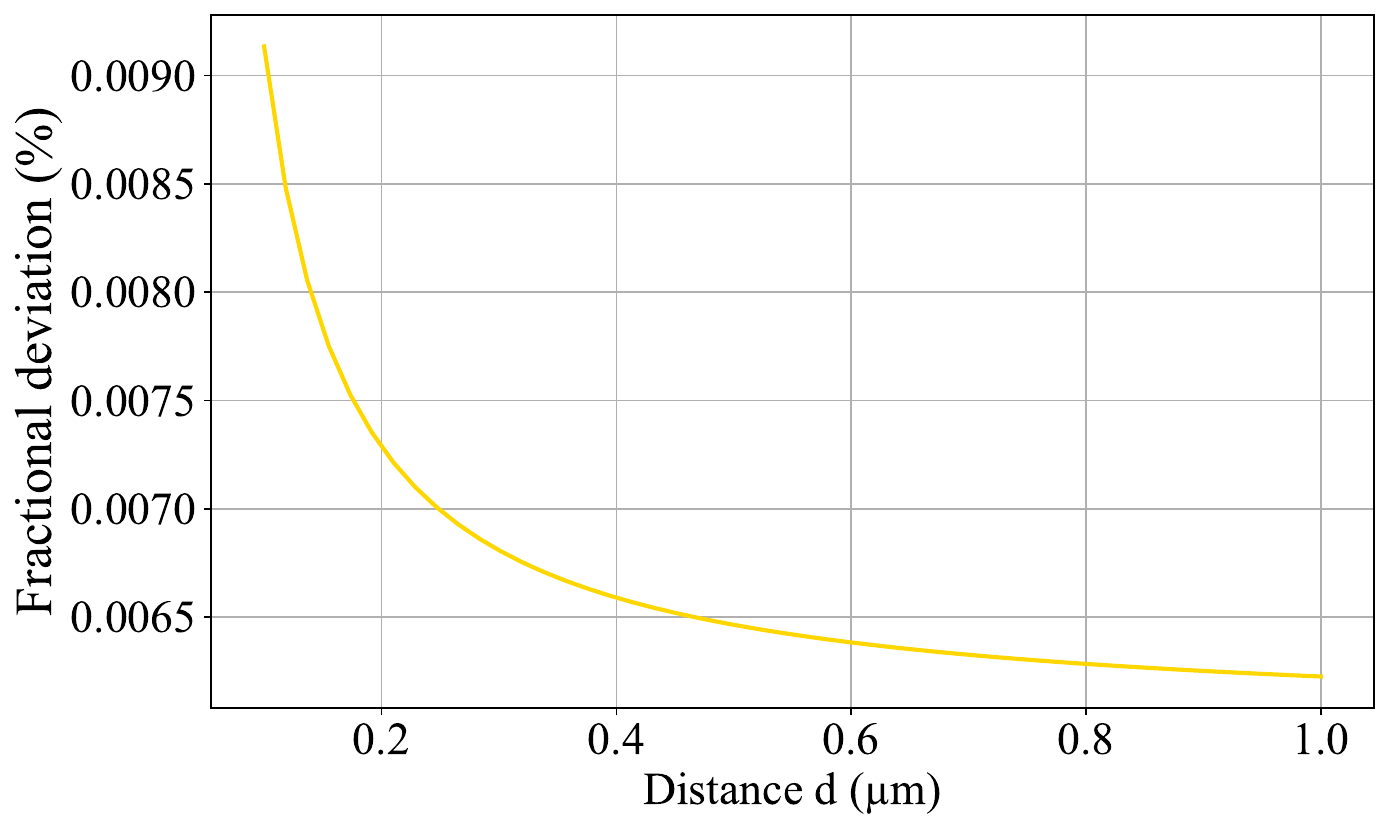}
    \caption{}
    \label{fig:graph_error_a}
    \end{subfigure}
    \hfill
    \begin{subfigure}[b]{0.48\textwidth}
     \includegraphics[width=1\textwidth]{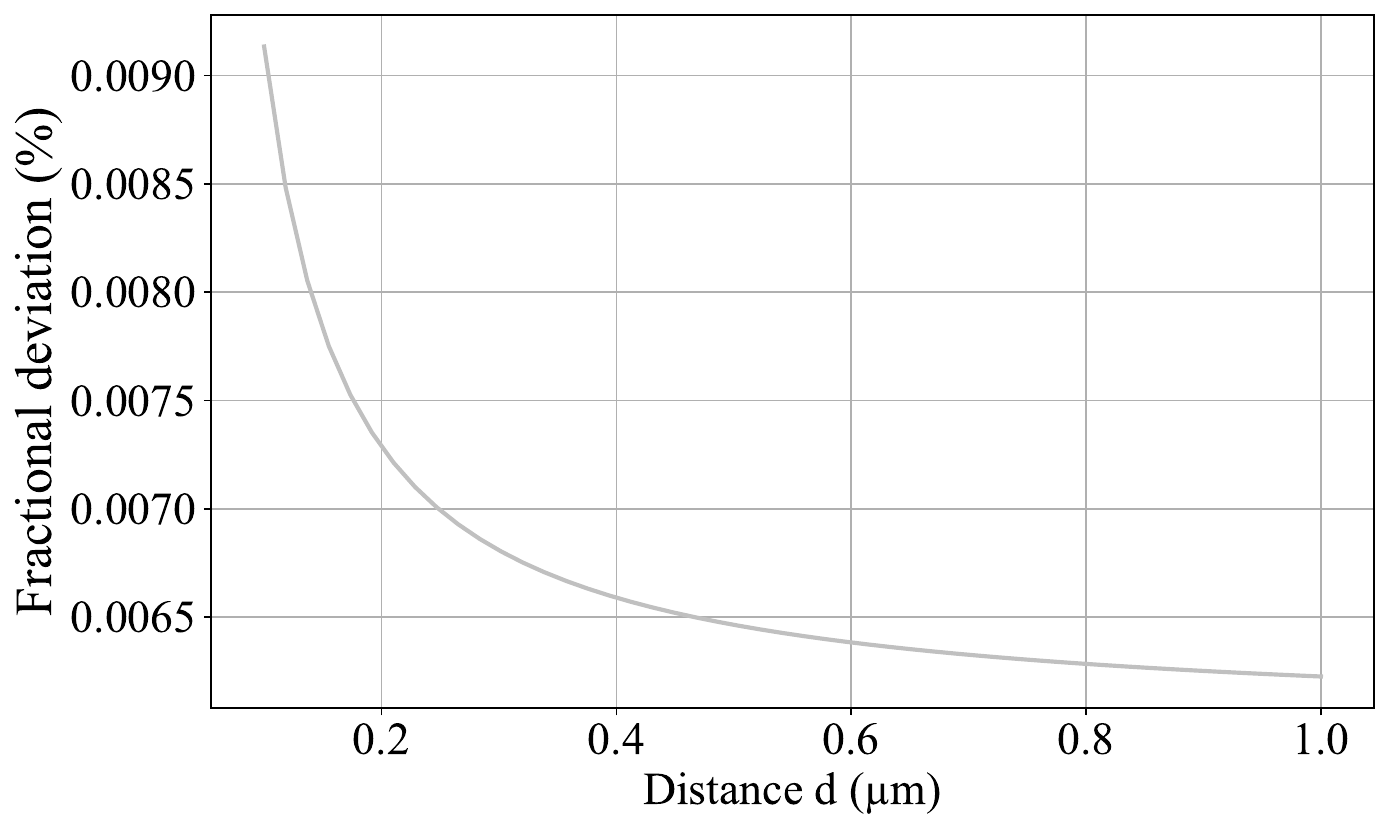}
    \caption{}
    \label{fig:graph_error_b}
    \end{subfigure}
\caption{\raggedright {Fractional deviation between the leading-order PFA and the PFA with NTLO correction as a function of separation distance for}{ (a) for Au }{(b) for Ag.}}
    \label{fig:error_comparison}
\end{figure}

Different geometries could result in a higher fractional deviation value that accounts for curvature-dependent effects. Regarding the arc-to-plate geometry in this study and its quasi-parallel nature, the PFA with perturbation parameter $\epsilon = 0.1$ serves as a good approximation, while PFA with NTLO further precises it. However, we consider the NTLO correction effects as negligible. The thickness range of the PFA  perturbation parameter $\epsilon = 0.1$ can be observed in table~\ref{fig:table_PFA_perturbation}.  Detailed data for all values of the distance range $d \in [0.1 - 1]\mu m$ are shown in Fig.~\ref{fig:graph_error_a} and Fig.~\ref{fig:graph_error_b}. Since the deviation can be expressed as a ratio of energies that factors out the material-dependent constants, it is identical for both silver and gold. However, it should be noted that violating the PFA condition $d<<R$ can result in significantly higher fractional deviations.

\section{Discussion}

\subsection{Applications}

The Casimir effect has applications in MEMS, and under different circumstances, it can benefit the system or cause difficulties. In some scenarios, the Casimir effect is regarded as a drawback, that is, the components of the microdevice experience stiction during the fabrication process. Moreover, stiction is generally observed in cantilever MEMS switches. Although this study investigates the maximum thickness for curvature reversal, the same calculations are useful for estimating the value for which the nanomembrane will not undergo curvature reversal, which occurs when the energy required to bend the arc is greater than the energy due to the Casimir force. That is, stiction prevention will occur when the thickness lies in the upper range of the estimated data; the thickness value should be greater than the values obtained in this study for the specific distance observed. Choosing the thickness value of any surface that is influenced by the Casimir force is crucial to obtaining certainty that the MEMS components will not collapse; in other words, it will not result in stiction of the components. The application of the estimated thickness values for a required MEMS provides direction for more stable nanomembrane designs for metrology and quantum devices. 

Although the Casimir pull-in is a disadvantage of the Casimir effect, the Casimir force has practical applications in MEMS, as it is a source of energy at the micro- and sub-micron levels. Attempts have been made to tune the Casimir force to obtain control over its magnitude, with applications such as shifting the force from attractive to repulsive. In addition to the method involving tuning the Casimir force by dielectric control as observed in detail in Section 1.3 of this paper (\citep{munday_measured_2009},  \citep{zhao_stable_2019}), \citep{passante_quantum_2025} reviews tuning the Casimir force by external magnetic fields. This suggests that the restoring force can be obtained by using tuning methods.
Enabling the use of tuning as a restoration mechanism creates the possibility of incorporating cyclic use in MEMS where the Casimir force performs work, and a restoring force brings the system to the initial position. Such cyclic interactions have applications in casimir-driven actuators. 
Designing a bridge-type MEMS switch with the required thickness and active Casimir force tuning would enable the use of the Casimir effect as an energy source.
 This is an alternative to cantilever switch design, which uses quantum energy and overcomes stiction. These switches find their best applications in vacuum because the Casimir force is the strongest. The required environment enables their use in MEMS for space applications. Harnessing the Casimir effect as an energy source allows for a lower power input; consequently, the system could operate longer for the equivalent power input. As reported by \citep{pacheco_design_2000}, MEMS switch solid-state equivalents require higher drive power.

Curvature reversal allows the design of a micropump. The micropump is composed of the arc-plate geometry, where the Casimir force drives the motion, making it a non-mechanical pump, as it uses Casimir energy to produce kinetic energy. Returning to the initial position of curvature, that is, reversing the curvature of the nanomembrane from convex to concave with respect to the plate can be achieved using the proposed restoring forces acquired from Casimir force tuning. Micropumps have many applications, some of which include cooling in nanoelectronics and fuel delivery in nanorobotics. Moreover, micropumps have applications in the biomedical sphere for drug delivery, as in \citep{lo_passive_2009}. All three may benefit from the incorporation of the Casimir effect as either an energy or pressure source.

Casimir forces can become relevant in critical regimes such as ultra-thin membrane designs, near-contact operation, or failure-limiting scenarios including pull-in and stiction. The results of this study therefore do not imply dominance of Casimir effects during normal MEMS operation, but rather identify conservative bounds under which Casimir-induced deformation may become significant.

\subsection{Future developments}

Further developments include an experimental evaluation of the Casimir energy obtained by the arc-to-plate geometry for both gold and silver. In this study, the thickness of the 2D geometry was evaluated. That is, the 2D geometry $x = d+ \sqrt{R^2-y^2}$ is considered, whereas in 3D, a spherical rather than a cylindrical arc geometry can be evaluated. Estimating the thickness of the 3D equivalent of the geometry will account for the impact of the surface of the nanomembrane on the acquired force. Furthermore, evaluating the bending energy will provide data for various materials, which is beneficial for applications in MEMS because of the varying material properties. Moreover, further restoring force mechanisms can be incorporated to create a reliable Casimir-driven MEMS that requires minimal power input. Finally, given that this work presents purely theoretical interpretations, experimental validation is left to future research.

\section{Conclusions}

The Casimir effect was observed for the arc-plate geometry. Specifically, a theoretical estimate of the Casimir energy required to bend a nanomembrane from the initial position, which is concave to the plate, to the final position, which is convex to the plate, was conducted. The Casimir energy is calculated using the PFA method with NTLO correction for an array of 1000 distances in the range $d \in [0.1 - 1 ]\mu m$; where the radius of the arc is $100 \mu m$ in order to satisfy the PFA condition, i.e. $d<<R$. The maximum thickness was computed by comparing the Casimir energy to the estimated bending energy of the nanomembranes for gold and silver. The maximum thickness to be able to perform curvature reversal is obtained in the range $9.636$ nm \textendash $ 7.817 \times 10^{-1}$ nm for silver and $9.532$ nm \textendash $7.732 \times 10^{-1}$ nm for gold. The precision of the NTLO correction was evaluated in comparison with PFA with a perturbation parameter $\epsilon = 0.1$.  Applications in MEMS are discussed, including methods for overcoming stiction in MEMS cantilever switches and Casimir-driven devices such as micropumps. 
Future work could focus on coupling the present analytical framework with finite-element simulations to investigate Casimir-induced deformation and stability under realistic boundary conditions.
This work provides a theoretical framework for Casimir forces in nanofilms, which could serve as a basis for future numerical simulations and experimental studies.

\vspace{6pt} 

\section{Acknowledgements}{Conceptualization, A.A.; Methodology, A.A.; Software, A.A.; Validation, A.A.; Formal analysis, A.A.; Investigation, A.A.; Resources, A.A.; Data curation, A.A.; Writing---original draft preparation, A.A.; Writing---review and editing, J.V. and A.A.; Visualization, A.A.; Supervision, J.V. and A.A.; Project administration, A.A.; funding acquisition, A.A. All authors have read and agreed to the published version of the manuscript. This research received no external funding. All the data generated and analyzed for this study are included in this article. The source code used to obtain the data is available at \url{https://github.com/alexandrovaannamaria/Casimir-Arc-Plate-Geometry-Thickness-Constraints-/tree/main}; doi: \url{https://zenodo.org/records/17246424}. The authors declare no conflicts of interest.
} 

\section{Funding}
This work was funded by the Vicerrectorado de Investigación (VRI) at the Pontificia Universidad Católica del Perú through the Dirección de Fomento de la Investigación (Project ID: PI1287).

\section*{}


\begin{thebibliography}{99}
\bibliographystyle{unsrt}


\bibitem{casimir_influence_1948}
Casimir, H. B. G.; Polder, D. The Influence of Retardation on the London-van Der Waals Forces. \textit{Phys. Rev.} \textbf{1948}, \textit{73}, 360--372. https://doi.org/10.1103/PhysRev.73.360.

\bibitem{jaffe_casimir_2005}
Jaffe, R. L. The Casimir Effect and the Quantum Vacuum. \textit{Phys. Rev. D} \textbf{2005}, \textit{72}, 021301. https://doi.org/10.1103/PhysRevD.72.021301.

\bibitem{lamoreaux_casimir_2005}
Lamoreaux, S. K. The Casimir Force: Background, Experiments, and Applications. \textit{Rep. Prog. Phys.} \textbf{2005}, \textit{68}, 201--236. https://doi.org/10.1088/0034-4885/68/1/R04.

\bibitem{milonni_quantum_1994}
Milonni, P. \textit{The Quantum Vacuum}; Elsevier, 1994. https://doi.org/10.1016/C2009-0-21295-5.

\bibitem{lamoreaux_demonstration_1997}
Lamoreaux, S. K. Demonstration of the Casimir Force in the 0.6 to 6 $\mu$m Range. \textit{Phys. Rev. Lett.} \textbf{1997}, \textit{78}, 5--8. https://doi.org/10.1103/PhysRevLett.78.5.

\bibitem{garrett_measurement_2018}
Garrett, J. L.; Somers, D. A. T.; Munday, J. N. Measurement of the Casimir Force between Two Spheres. \textit{Phys. Rev. Lett.} \textbf{2018}, \textit{120}, 040401. https://doi.org/10.1103/PhysRevLett.120.040401.

\bibitem{emig_casimir_2006}
Emig, T.; Jaffe, R. L.; Kardar, M.; Scardicchio, A. Casimir Interaction between a Plate and a Cylinder. \textit{Phys. Rev. Lett.} \textbf{2006}, \textit{96}, 080403. https://doi.org/10.1103/PhysRevLett.96.080403.

\bibitem{fosco_proximity_2011}
Fosco, C. D.; Lombardo, F. C.; Mazzitelli, F. D. Proximity Force Approximation for the Casimir Energy as a Derivative Expansion. \textit{Phys. Rev. D} \textbf{2011}, \textit{84}, 105031. https://doi.org/10.1103/PhysRevD.84.105031.

\bibitem{fosco_casimir_2024}
Fosco, C. D.; Lombardo, F. C.; Mazzitelli, F. D. Casimir Physics beyond the Proximity Force Approximation: The Derivative Expansion. \textit{Physics} \textbf{2024}, \textit{6}, 290--316. https://doi.org/10.3390/physics6010020.

\bibitem{stange_building_2019}
Stange, A.; Imboden, M.; Javor, J.; Barrett, L. K.; Bishop, D. J. Building a Casimir Metrology Platform with a Commercial MEMS Sensor. \textit{Microsyst. Nanoeng.} \textbf{2019}, \textit{5}, 14. https://doi.org/10.1038/s41378-019-0054-5.

\bibitem{javor_analysis_2021}
Javor, J.; Yao, Z.; Imboden, M.; Campbell, D. K.; Bishop, D. J. Analysis of a Casimir-Driven Parametric Amplifier with Resilience to Casimir Pull-in for MEMS Single-Point Magnetic Gradiometry. \textit{Microsyst. Nanoeng.} \textbf{2021}, \textit{7}, 73. 
https://doi.org/10.1038/s41378-021-00289-4.

\bibitem{javor_zeptometer_2022}
Javor, J.; Imboden, M.; Stange, A.; Yao, Z.; Campbell, D. K.; Bishop, D. J. Zeptometer Metrology Using the Casimir Effect. \textit{J. Low Temp. Phys.} \textbf{2022}, \textit{208}, 147--159. https://doi.org/10.1007/s10909-021-02650-3.

\bibitem{busch_heisenbergs_2007}
Busch, P.; Heinonen, T.; Lahti, P. Heisenberg's Uncertainty Principle. \textit{Phys. Rep.} \textbf{2007}, \textit{452}, 155--176.  
https://doi.org/10.1016/j.physrep.2007.05.006

\bibitem{casimir_attraction_nodate}
Casimir, H. B. G. On the Attraction between Two Perfectly Conducting Plates. \textit{Proc. K. Ned. Akad. Wet.} \textbf{1948}, \textit{51}, 793--795.

\bibitem{jackson_classical_2009}
Jackson, J. D. \textit{Classical Electrodynamics}, 3rd ed.; Wiley: Hoboken, NY, 2009.

\bibitem{munday_measured_2009}
Munday, J. N.; Capasso, F.; Parsegian, V. A. Measured Long-Range Repulsive Casimir--Lifshitz Forces. \textit{Nature} \textbf{2009}, \textit{457}, 170--173. https://doi.org/10.1038/nature07610.

\bibitem{zhao_stable_2019}
Zhao, R.; Li, L.; Yang, S.; Bao, W.; Xia, Y.; Ashby, P.; Wang, Y.; Zhang, X. Stable Casimir Equilibria and Quantum Trapping. \textit{Science} \textbf{2019}, \textit{364}, 984--987. https://doi.org/10.1126/science.aax0916.

\bibitem{rodriguez_repulsive_2008}
Rodriguez, A. W.; Joannopoulos, J. D.; Johnson, S. G. Repulsive and Attractive Casimir Forces in a Glide-Symmetric Geometry. \textit{Phys. Rev. A} \textbf{2008}, \textit{77}, 062107. https://doi.org/10.1103/PhysRevA.77.062107.

\bibitem{ederth_template-stripped_2000}
Ederth, T. Template-Stripped Gold Surfaces with 0.4 nm RMS Roughness Suitable for Force Measurements: Application to the Casimir Force in the 20--100 nm Range. \textit{Phys. Rev. A} \textbf{2000}, \textit{62}, 062104. https://doi.org/10.1103/PhysRevA.62.062104.

\bibitem{banishev_modifying_2012}
Banishev, A. A.; Chang, C.-C.; Castillo-Garza, R.; Klimchitskaya, G. L.; Mostepanenko, V. M.; Mohideen, U. Modifying the Casimir Force between Indium Tin Oxide Film and Au Sphere. \textit{Phys. Rev. B} \textbf{2012}, \textit{85}, 045436. https://doi.org/10.1103/PhysRevB.85.045436.

\bibitem{chen_investigation_2005}
Chen, F.; Mohideen, U.; Klimchitskaya, G. L.; Mostepanenko, V. M. Investigation of the Casimir Force between Metal and Semiconductor Test Bodies. \textit{Phys. Rev. A} \textbf{2005}, \textit{72}, 020101. https://doi.org/10.1103/PhysRevA.72.020101.

\bibitem{chaparro_oxidation_2023}
Chaparro, D.; Goudeli, E. Oxidation Rate and Crystallinity Dynamics of Silver Nanoparticles at High Temperatures. \textit{J. Phys. Chem. C} \textbf{2023}, \textit{127}, 13389--13397. https://doi.org/10.1021/acs.jpcc.3c03163.

\bibitem{chaparro_shape-dependent_2024}
Chaparro, D.; Goudeli, E. Shape-Dependent Oxidation Rates of Nano-Structured Silver Particles. \textit{J. Chem. Phys.} \textbf{2024}, \textit{161}, 124704. https://doi.org/10.1063/5.0227329.

\bibitem{mohideen_precision_1998}
Mohideen, U.; Roy, A. Precision Measurement of the Casimir Force from 0.1 to 0.9 $\mu$m. \textit{Phys. Rev. Lett.} \textbf{1998}, \textit{81}, 4549--4552. https://doi.org/10.1103/PhysRevLett.81.4549.


\bibitem{klimchitskaya_casimir_2024}
Klimchitskaya, G. L.; Korotkov, A. S.; Loboda, V. V.; Mostepanenko, V. M. Role of the Casimir Force in Micro- and Nanoelectromechanical Pressure Sensors. \textit{EPL } \textbf{2024}, \textit{146} (6), 66004. https://doi.org/10.1209/0295-5075/ad4fbb


\bibitem{gies_casimir_2006}
Gies, H.; Klingmüller, K. Casimir Edge Effects. \textit{Phys. Rev. Lett.} \textbf{2006}, \textit{97}, 220405. https://doi.org/10.1103/PhysRevLett.97.220405.


\bibitem{bimonte_measurement_2021}
Bimonte, G.; Spreng, B.; Maia Neto, P. A.; Ingold, G.-L.; Klimchitskaya, G. L.; Mostepanenko, V. M.; Decca, R. S. Measurement of the Casimir Force between 0.2 and 8 $\mu$m: Experimental Procedures and Comparison with Theory. \textit{Universe} \textbf{2021}, \textit{7}, 93. https://doi.org/10.3390/universe7040093.


\bibitem{emam_review_2022}
Emam, S.; Lacarbonara, W. A Review on Buckling and Postbuckling of Thin Elastic Beams. \textit{Eur. J. Mech. A/Solids} \textbf{2022}, \textit{92}, 104449. https://doi.org/10.1016/j.euromechsol.2021.104449


\bibitem{wierzbicki_tomasz_2081j16230j_nodate}
Wierzbicki, T. 2.081J/16.230J Plates and Shells; Massachusetts Institute of Technology, 2006.
\url{https://ocw.mit.edu/courses/2-081j-plates-and-shells-spring-2007/1bd32d641a2b41ca518591d981405393_lecturenote.pdf}

\bibitem{noel_review_2016}
Noel, J. G. Review of the Properties of Gold Material for MEMS Membrane Applications. \textit{IET Circuits Devices Syst.} \textbf{2016}, \textit{10}, 156--161. https://doi.org/10.1049/iet-cds.2015.0094.

\bibitem{ruud_elastic_1991}
Ruud, J.; Josell, D.; Greer, A. L.; Spaepen, F. The Elastic Moduli of Silver Thin Films Measured with a New Microtensile Tester. \textit{MRS Proc.} \textbf{1991}, \textit{239}, 239. https://doi.org/10.1557/PROC-239-239.

\bibitem{passante_quantum_2025}
Passante, R.; Rizzuto, L.; Schall, P.; Marino, E. Quantum and Critical Casimir Effects: Bridging Fluctuation Physics and Nanotechnology. \textit{Nanoscale} \textbf{2025}, \textit{17}, 13982--13997. https://doi.org/10.1039/D5NR01288K.

\bibitem{pacheco_design_2000}
Pacheco, S. P.; Katehi, L. P. B.; Nguyen, C. T.-C. Design of Low Actuation Voltage RF MEMS Switch. In \textit{2000 IEEE MTT-S International Microwave Symposium Digest}; IEEE: Boston, MA, USA, 2000; Vol. 1, pp 165--168. https://doi.org/10.1109/MWSYM.2000.860921.

\bibitem{lo_passive_2009}
Lo, R.; Li, P.-Y.; Saati, S.; Agrawal, R. N.; Humayun, M. S.; Meng, E. A Passive MEMS Drug Delivery Pump for Treatment of Ocular Diseases. \textit{Biomed. Microdevices} \textbf{2009}, \textit{11}, 959--970. 
\end{thebibliography}
\end{document}